\documentstyle[prl,aps,preprint,tighten,floats]{revtex}

\def\beq{\begin{equation} }
\def\eeq{\end{equation} }
\def\beqn{\begin{eqnarray} }
\def\eeqn{\end{eqnarray} }
\def\be{\begin{equation} }
\def\ee{\end{equation} }
\def\ben{\begin{eqnarray} }
\def\een{\end{eqnarray} }
\def\bear{\begin{eqnarray}}
\def\eear{\end{eqnarray}}

\def\hb#1{\hbox to 1.0truecm{#1\hfill}}
\def\section#1{\vspace{1\baselineskip}
		\noindent{\bf #1}\\
		}
\def\subsection#1{\vspace{\baselineskip}
		\noindent{\bf #1}\\
		}

\def\Tr{{\rm Tr}}

\def\vev#1{\langle #1\rangle}

\def\MK{M }

\def\MP{M_{\rm Pl} }

\def\MS{M_{\rm string} }

\def\MS{M_{\rm string}}

\def\MZ{M_{\rm Z} }

\def\mur{\mu_{rad}}

\def\MZ{M_{\rm Z}}

\fnsymbol{footnote}

\begin{document}
\begin{titlepage}
\samepage{
\setcounter{page}{1}
\rightline{CTP-TAMU-40/98}
\rightline{UPR-795-T}
\rightline{\tt hep-ph/9812262}
\rightline{December 1998}
\vspace{1.0cm}
\begin{center}
{\Large \bf Quark Masses and Flat Directions in String Models\\}
\vspace{.75cm} 
\vfill {\large
	Gerald B. Cleaver\footnote{gcleaver@harc.edu}}\\
\vspace{.40in}
{\it    Department of Physics and Astronomy\\
        The University Of Pennsylvania\\
        Philadelphia, Pennsylvania  19104-6396\\}
\vspace{.15in}
{\it    Center for Theoretical Physics\\
        Department of Physics\\ 
        Texas A\&M University,\\
        College Station, TX 77843\\}
\vspace{.07in}
        and\\
\vspace{.07in}
{\it    Astro Particle Physics Group\\
        Houston Advanced Research Center (HARC)\\
        The Mitchell Campus,Woodlands, TX 77381\\}
\vspace{.01in}
\end{center}
\begin{abstract}
I discuss a method for producing a quasi--realistic 
inter--generational quark mass hierarchy in string models. 
This approach involves non--Abelian singlet states developing 
intermediate scale vacuum expectation values. I summarize 
recent investigations into string model realization of this. 
\end{abstract}
\vskip 0.5truecm
\centerline{\it Talk presented at QCD98, Paris, France, 1--6 June 1998.}
\smallskip}
\vfill
\end{titlepage}

\setcounter{footnote}{0}

\section{I. Quark Mass Hierarchy}

The experimental measurement a few years ago of the top quark mass  
\cite{prd96} eliminated the last unknown among     
the physical masses of {\it all} three generations of up--, down--, 
and electron--like particles.
The masses of these particles are displayed in Table \ref{masstable1},
where I have expressed all masses in top--quark mass units.
To first order, the inter--generational mass ratio for up--type quarks is 
$10^{-5}:10^{-3}:1$, 
and for down--type is $10^{-5}:10^{-3}:10^{-2}$.  

In minimal supersymmetric standard model (MSSM) physics, 
quarks (and their supersymmetric partners) 
gain mass through superpotential couplings  
to Higgs bosons $H_1$ and $H_2$,\footnote{$\hat{X}$ 
denotes a generic superfield and $X$ its bosonic component.}
\beq 
W_{u_i}\sim \lambda_{u_i} \hat{H}_2 \hat{Q_i} \hat{U}^c_i\, ;\quad 
W_{d_i}\sim \lambda_{d_i} \hat{H}_1 \hat{Q_i} \hat{D}^c_i\, ;\quad 
\label{nonrenql}
\eeq 
where $i$ is the generation number. Effective mass terms appear
when the Higgs acquire a typical soft supersymmetry breaking scale 
vacuum expectation value (VEV)
$\vev{H_{1,2}} \sim  m_{soft}= {\cal O} (M_Z)$.\footnote{Similarly, 
the three generations of electron--type leptons gain mass via superpotential
terms $W_{e_i}\sim \lambda_{e_i} \hat{H}_1 \hat{L_i} \hat{E}^c_i$.}

Inter--generational mass ratios can be induced when the associated
first and second generation superpotential terms contain 
effective couplings $\lambda$ that include non--renormalizable
suppression factors,  
\beq
\lambda_{(u,d,e)_i} \sim (\frac{\vev{S}}{\MP})^{P'_i},
\quad {\rm ~for~} i=1,2,
\label{pval}
\eeq 
where
$S$ is a non--Abelian singlet, $\MP$ is the Planck scale (which is replaced 
by the $\MS$ for string models), and $P$ is a positive integer.
VEVs only slightly below the Planck/string scale 
(often resulting from a $U(1)$ anomaly cancellation) imply large values 
of $P'_{1,2}$ for $10^{-5}$ and $10^{-3}$ suppression 
factors. In contrast, intermediate scale VEVs
(between $\MZ$ and $\MS$) require far lower values for $P'_i$.
In a series of recent papers  
intermediate scales have been explored \cite{CCEEL1,CCEEL2}
and their realization in actual models has been 
investigated  \cite{CCEELW,GC1}.   
This investigation has been the product of a
fruitful collaboration with M. Cveti\v c, J. Espinosa, L. Everett, 
P. Langacker, \& J. Wang at the University of 
Pennsylvania.\footnote{In addition to this mass ratio study 
from the perturbative string theory perspective,
Katsumi Tanaka of the Ohio State University 
and I have been investigating 
quark mass ratios from the non--perturbative approach of 
Seiberg--Witten duality \cite{gckt}.} 
In the following section, I show how intermediate scales can occur and
how, in theory, they could produce 
an inter--generational $10^{-5} : 10^{-3}: 1$ up--quark mass ratio,
and a corresponding $10^{-5}:10^{-3}:10^{-2}$ down--quark mass ratio.  

\hfill\vfill\eject

\section{II. Theoretical Quark Mass Hierarchy from String Models}

One method by which an intermediate scale VEV $\vev{S}$ 
can generate non--renormaliz-able suppression factors 
involves extending the SM gauge group by an additional non--anomalous 
$U(1)'$. This approach requires (at least)
two SM singlets $S_1$ and $S_2$, 
carrying respective $U(1)'$ charges $Q'_1$ and $Q'_2$.
$D$--flatness for the non--anomalous $U(1)'$,  
\beq
\vev{D}_{U(1)'} \equiv Q'_{1}|\vev{S_1}|^2 + Q'_{2}|\vev{S_2}|^2 = 0,
\label{dif}
\eeq
necessitates that $Q'_{1}$ and $Q'_{2}$ be of opposite sign.
Together the VEVs of $S_1$ and $S_2$ form a $D$--flat scalar field 
direction $S$ defined by,
\beq
\langle S_1\rangle=\cos\alpha_Q\langle S\rangle,\;\;\;\;
\langle S_2\rangle=\sin\alpha_Q\langle S\rangle,\quad
{\rm where}\quad
\tan^2\alpha_Q\equiv \frac{|Q_1|}{|Q_2|}.
\label{flatdir}
\eeq

The $F$--flatness constraints  
\beq
\vev{F_{S_p}} \equiv \vev{\frac{\partial W}{\partial S_p }} = 0,\,\, p= 1,2;
\quad {\rm and}\quad \vev{W}=0,
\label{ff}
\eeq
imply that the $D$--flat direction
$S= S_1 \cos\alpha_Q  + S_2 \sin\alpha_Q$ is also a
{\it renormalizable} $F$--flat direction if (as I assume hereon)
$\hat{S}_1$ and $\hat{S}_2$ do not couple among themselves
in the renormalizable superpotential. 
 
Consider the real component of this flat direction, 
$s=\sqrt{2} {\mathrm Re} S = s_1 \cos\alpha_Q  + s_2 \sin\alpha_Q$. 
This scalar's renormalization group equation (RGE) running mass is,
\beq
m^2=m_1^2(\mu) \cos^2\alpha_Q+m_2^2(\mu)\sin^2\alpha_Q=\left(
\frac{m_1^2}{|Q_1|}+\frac{m_2^2}{|Q_2|}\right)
\frac{|Q_1Q_2|}{|Q_1|+|Q_2|},
\label{msum}
\eeq
which generates a potential  
\beq
V(s)=\frac{1}{2 }m(\mu= s)^2 s^2.
\label{mpot}
\eeq

I will assume that $m^2$ is positive at the string scale
and of order $m^2_{\rm soft} \sim {\cal O}({\MZ}^2)$
($m^2_o$ if universality is assumed).
However, through RGE running, $m^2$ can
be driven negative (with electroweak (EW) scale magnitude) 
by large Yukawa couplings  
(i) of $S_1$ to exotic triplets, $W=h\hat{D}_1\hat{D}_2\hat{S}_1$;
(ii) of $S_1$ to exotic doublets and of $S_2$ to exotic triplets, 
    $W=h_D\hat{D}_1\hat{D}_2\hat{S}_1+h_L\hat{L}_1\hat{L}_2\hat{S}_2$; or
(iii) of $S_1$ to varying numbers of additional SM singlets
$W=h\sum_{i=1}^{N_p}\hat{S}_{ai}\hat{S}_{bi}\hat{S}_1$ \cite{CCEEL1}. 
$m(\mu= s)^2$ can turn negative anywhere between a scale of
$\mur= 10^4$ GeV and $\mur= 10^{17}$ GeV (slightly below the string scale) 
for various choices of the supersymmetry breaking parameters
$A^0$ (the universal Planck scale soft trilinear coupling) and
$M_{1/2}$ (the universal Planck scale gaugino mass).\footnote{The
standard universal scalar EW soft mass--squared parameter $m_0$
has a simple normalizing effect, with $A^0/m_0$ and $M_{1/2}/m_0$
being the actual relevant parameters.}   
  
When $m^2$ runs negative,
a minimum of the potential develops along the flat direction
and $S$ gains a non--zero VEV. 
In the case of only a mass term and no Yukawa contribution to $V(s)$,
minimizing the potential
\begin{equation}
\frac{dV}{ds}=\left.\left(m^2+\frac{1}{2}\beta_{m^2}\right)\right|_{\mu=s}s=0,
\label{vdeq}
\end{equation}
(where $\beta_{m^2}=\mu\frac{dm^2}{d\mu}$) 
shows that the VEV $\langle s\rangle$ is determined by 
\begin{equation}
m^2(\mu=\langle s\rangle)=-\frac{1}{2}\beta_{m^2}.
\label{m2eq}
\end{equation}
Eqs.\ (\ref{vdeq},\ref{m2eq})
are satisfied very close to the scale $\mu_{RAD}$ at which $m^2$ crosses
zero.  $\mu_{RAD}$ is fixed by the renormalization group evolution of
parameters from $\MS$ down to the EW scale and will lie
at some intermediate scale. 

Location of the potential minimum can also be effected by non--renormalizable
self--interaction terms,
\begin{equation}
W_{\rm NR}=\left(\frac{\alpha_{K}}{\MP}\right)^{K}\hat{S}^{3+K},
\label{nrsup}
\end{equation}
where $K=1,2...$ and $\alpha_{K}$ are coefficients. 
Such non--renormalizable operators (NRO's) 
lift the flat direction 
(by breaking $F$--flatness)
for sufficiently large values of $s$. 

The general form of the  potential, $V(X_p)$, for the scalar components 
$X_p$ of corresponding supermultiplets $\hat{X_p}$ is
\beqn
V(X_p) &=& V_{soft\,\, susy} + \sum_p \mid  \frac{\partial W}{\partial \hat{X}_p} \mid^2 
           +\frac{1}{2} g_{\alpha}^2\sum_\alpha  
            \mid  \sum_p Q_p^{\alpha} | X_p \mid^2 \mid^2 
\label{vsnroa}\\
          &=& V_{soft\,\,susy} + \sum_p \mid F_p \mid^2 \phantom{0}\phantom{0}    
           +\frac{1}{2} g_{\alpha}^2\sum_\alpha \mid D_{\alpha}\mid^2 
\label{vsnrob}
\eeqn
Thus, NRO contributions transform $V(s)$ in eq.\ (\ref{mpot}) into 
\beq
V(s)= \frac{1}{2}m^2s^2+\frac{1}{2(K+2)}
\left(\frac{s^{2+K}}{{\cal M}^K}\right)^2,
\label{vsnro}
\eeq
where ${\cal M} = {\cal C}_K \MP/ \alpha_K$, with
${\cal C}_K = [ 2^{K+1}/((K+2)(K+3)^2)]^{1/(2K)}$.

Even when an NRO is present,
the running mass effect still dominates in determining $\vev{s}$
if $\mu_{RAD}\ll 10^{12}$ GeV.  
However, an NRO is the controlling factor when
$\mu_{RAD}\gg 10^{12}$ GeV. In the latter case, we find that
\begin{equation}
\label{veveq}
\langle s\rangle=
\left[\sqrt{(-m^2)}{\cal M}^K \right]^\frac{1}{K+1}
= \mu_K\sim (m_{soft}{\cal M}^K)^\frac{1}{K+1},
\end{equation}
where $m_{soft}={\cal O} (|m|)={\cal O} (M_Z)$ is a typical soft supersymmetry
breaking scale.
While $\vev{s}$ is an intermediate scale VEV, the mass $M_S$ of
the physical field $s$ is still on the order of the soft SUSY breaking scale:
For running mass domination,
\begin{equation}
M_S^2\equiv\left.\frac{d^2V}{ds^2}\right|_{s=\langle s \rangle}=
\left.\left(\beta_{m^2}+\frac{1}{2}\mu\frac{d}{d\mu}\beta_{m^2}\right)
\right|_{\mu=\langle s\rangle}\simeq \beta_{m^2}\sim\frac{m^2_{soft}}{16\pi^2},
\end{equation}
while in the NRO--controlled case,
\begin{equation}
M_S^2=2(K+1)(-m^2)\sim m^2_{soft}.
\end{equation}

What powers $P'_i$ in (\ref{pval})
for first and second generation suppression factors  
in an NRO--dominated model could produce 
an up--type quark mass ratio of order $10^{-5}:10^{-3}:1$?
From eq.\ (\ref{veveq}), we see the suppression factors become
\beq
\left(\frac{m_{soft}}{\MK}\right)^{\frac{P'_i}{K+1}},
\label{prat}
\eeq
where the coefficient $\alpha_K$ has been absorbed into 
the definition of the mass scale $M\equiv {\cal M}/{\cal C}_K$. 
The mass suppression factors for specific $P'$ (in the range $0$ to $5$)
and $K$ (in the range of $1$ to $7)$ are given in Table \ref{masstable2}.
From this table we find that the choices $P'_1= 2$ and $P'_2= 1$ in tandem with
$K=5$ or $K=6$ (for the self--interaction terms of $S$) 
can, indeed, reproduce the required mass ratio. These values of $K$
invoke an intermediate scale $\vev{S}$ around
$8\times 10^{14}$ GeV to $2\times 10^{15}$ GeV, 
 
The first and second generation down--quark suppression factors
can be similarly realized.
However, unless $tan\ \beta\equiv \frac{\vev{H_2}}{\vev{H_1}}\gg 1$,
the intra--generational mass ratio of
$10^{-2}: 10^{-2}: 1$ for $m_{\tau}$, $m_{b}$, and
$m_{t}$ is not realizable from a $K= 5$ or $6$ NRO singlet term. 
For $tan\ \beta \sim 1$, $m_{\tau}$, $m_{b}$ are 
too small to be associated with a renormalizable coupling ($P=0$) 
like that assumed for $m_{t}$, 
but are somewhat larger 
than predicted by $P= 1$ for $K=5$ or $K=6$. 
Instead, $m_b$ and $m_{\tau}$ might be 
associated with a different NRO involving the VEV of an entirely different 
singlet.
In that event, Table \ref{masstable2} suggests another 
flat direction $S'$ (formed from a second singlet pair $S_1'$ and $S_2'$), 
with a $K=7$ self--interaction NRO and $P'_3=1$ suppression factor for
$m_b$ and 
$m_{\tau}$.\footnote{An intermediate scale VEV $\vev{S}$ 
can also solve the $\mu$ problem through a superpotential term
$W_{\mu}\sim \hat{H_1} \hat{H_2}\hat{S}\left({\hat{S}\over \MK}\right)^{P_{\mu}}$.
With NRO--dominated $\langle S\rangle \sim (m_{soft} {\MK}^K)^{\frac{1}{K+1}}$,
the effective Higgs $\mu$--term takes the form,
$ \mu_{eff} \sim m_{soft} \left( {m_{soft}\over \MK} \right)^{\frac{P-K}{K+1}}$.
The phenomenologically preferred choice among this class of  
terms is clearly $P=K$: 
this yields a $K$--independent ${\mu}_{eff} \sim m_{soft}$.}

\section{III. Realization of Quark Mass Hierarchy in String Models}

In string models, one problem generally appears at the string
scale that must be resolved ``before'' possible intermediate 
scale flat direction VEVs can be investigated. 
That is, most four--dimensional quasi--realistic 
$SU(3)_C\times SU(2)_L\times U(1)_Y$  string models
contain an anomalous $U(1)_A$ (meaning $\Tr Q_A\ne 0$) \cite{KNCF}.
In fact, in a generic charge basis, a string model with an
Abelian anomaly may actually contain not just one, but
several anomalous $U(1)$ symmetries.
However, all anomalies can all be transferred into a single
$U(1)_A$ through the {\it unique} rotation  
\beq
         U(1)_{\rm A} \equiv c_A\sum_n \{\Tr Q_{n}\}U(1)_n,
\label{rotau1}
\eeq
with $c_A$ a normalization factor.
The remaining non--anomalous components of the original set of $\{U(1)_n\}$
may be rotated into a complete orthogonal basis $\{U(1)_a\}$.

The standard anomaly cancellation mechanism
\cite{DSW,ADS} breaks $U(1)_{\rm A}$ at the string scale, while simultaneously 
generating a FI $D$--term,
\beq
\xi\equiv \frac{g^2_s M_P^2}{192\pi^2}\Tr Q_A\, ,
\label{fid}
\eeq
where $g_{s}$ is the string coupling
and $M_P$ is the reduced Planck mass, 
$M_P\equiv M_{Planck}/\sqrt{8 \pi}\approx 2.4\times 10^{18}$. 
The FI $D$--term breaks spacetime supersymmetry unless it is cancelled by 
appropriate VEVs $\vev{X_p}$
of scalars $X_p$ that carry non--zero anomalous charge,
\beq
\vev{D}_{\rm A} \equiv \sum_j Q^{(A)}_j |\vev{X_p}|^2 + 
\xi
= 0\,\, .
\label{anomd}
\eeq
Generalizations of $D$-- and $F$--constraints for $S_{1,2}$ 
(i.e., of eqs.\  (\ref{dif}) and (\ref{ff}))
are imposed on possible VEV directions $\{\vev{X_p}\}$:  
$\vev{D}_{a} \equiv \sum_{p} Q^{(a)}_{p}|\vev{X_p}|^2 = 0$ and
$\vev{F_{p}} \equiv \vev{\frac{\partial W}{\partial \hat{X}_p}} = 0;
 \,\, \vev{W}  =0$.

My colleagues and I at Penn have developed 
methods for systematically determining \cite{dfset,GCM} 
and classifying \cite{CCEEL2} $D$-- and $F$--flat 
directions in string models.   
We have applied this process \cite{CCEELW,GC1} to the 
free fermionic three generation $SU(3)_C\times SU(2)_L\times U(1)_Y$ models 
(all of which contain an anomalous $U(1)_A$) 
introduced in refs.\ \cite{FNY}, \cite{AF} and \cite{CHLM}. 
For each model, we have determined the anomaly cancelling 
flat directions that preserve hypercharge,
only involve VEVs of non--Abelian singlet fields, and 
are $F$--flat to all orders in the non--renormalizable superpotential.
 
Flat directions in Model 5 of \cite{CHLM} have particularly received 
our attention \cite{CCEELW}.
In Model 5 we investigated the physics implications of 
various non--Abelian singlet flat directions. 
After breaking the anomalous $U(1)_A$, all of these flat directions 
left one or more additional $U(1)_a$ unbroken at the string scale.  
For each flat direction, the complete set of 
effective mass terms and effective trilinear superpotential
terms in the observable sector were computed {\it to all orders} in the
VEV's of the fields in the flat direction. 
The ``string selection--rules'' disallowed a large number of couplings 
otherwise allowed by gauge invariance, 
resulting in a massless spectrum with a large number
of exotics,\footnote{Recently it was shown \cite{CFN} 
that free fermionic construction can actually provide string models 
wherein {\it all} MSSM exotic states gain near--string--scale masses 
via flat direction VEVs, leaving only a string--generated MSSM in the 
observable sector below the string scale. 
The model of ref.\ \cite{FNY} was presented as the first known with these
properties.} 
which in most cases are excluded by experiment. 
This signified a generic flaw of these models.
Nevertheless, we found the resulting trilinear couplings of the
massless spectrum to possess a number of interesting features which
we analyzed for two representative flat directions. We investigated
the fermion texture;  baryon-- and lepton--number violating
couplings; $R$--parity breaking; non--canonical $\mu$ terms;
and the possibility of electroweak and intermediate scale symmetry breaking
scenarios for a $U(1)'\in \{ U(1)_a\}$.
The gauge coupling predictions were obtained in the electroweak scale case.
We found $t-b$ and $\tau-\mu$ fermion mass 
universality, with the string scale  Yukawa couplings $g$ and $g/\sqrt{2}$,
respectively. 
Fermion textures existed for certain flat directions, 
{\it but only in the down--quark sector.}
Lastly, we found
baryon-- and lepton-- number violating couplings that could 
trigger proton--decay, $N-{\bar N}$ oscillations,  leptoquark interactions 
and  $R$--parity violation, leading to the absence of a stable LSP.

\section{IV. Comments}

I have discussed how intermediate VEVs hold 
the potential to yield quasi--realistic quark mass ratios.
Four--dimensional string models usually require 
cancellation of the FI $D$--term contribution from an anomalous $U(1)_A$
by near string scale VEVs.   
Since some non--anomalous $U(1)_a$ are simultaneously broken at 
the string scale by these VEVs, 
which $U(1)_a$ might be associated with intermediate scale VEVs
strongly depends on the particular set of (near) string--scale VEVs chosen.
As our investigations into flat directions have demonstrated,
various choices for flat VEV directions can drastically alter
low energy phenomenology.  
The textures of the quark mass matrices can be strongly effected by choice
of flat direction since 
textures are wrought by effective mass terms in the non--renormalizable
superpotential.
      
\section{Acknowledgements}

G.C. thanks the organizers of QCD '98 for a very enjoyable and educational   
conference.
\hfill\vfill\eject

\begin{table}
\vskip 2.0truecm
\begin{tabular}{ccccccccccc}
$m_u$
&$:$
&$m_c$
&$:$
&$m_t$
&$=$ 
&$3\times 10^{-5}$
&$:$
&$7\times 10^{-3}$
&$:$
&$1$
\\
\hline
$m_d$
&$:$
&$m_s$
&$:$
&$m_b$ 
&$=$
&$6\times 10^{-5}$
&$:$
&$1\times 10^{-3}$
&$:$
&$3\times 10^{-2}$
\\
\hline
$m_e$
&$:$
&$m_{\mu}$
&$:$
&$m_{\tau}$ 
&$=$
&$0.3\times 10^{-5}$
&$:$
&$0.6\times 10^{-3}$
&$:$
&$1\times 10^{-2}$
\\
\end{tabular}
\caption{Fermion mass ratios with the top quark mass normalized to $1$. 
The values of $u-$, $d-$, and $s$-quark masses used in the ratios
(with the $t$-quark mass normalized to $1$ from an assumed mass of $170$ GeV) 
are estimates
of the $\overline{\rm MS}$ scheme current-quark masses at a scale 
$\mu\approx 1$ GeV. 
The $c$- and $b$-quark masses are pole masses.} 
\label{masstable1}
\end{table}
\begin{table}
\begin{tabular}{c|c|c|c|c|c|c|c|c}
       ~& $P^{(')}$ & $K=1$ & $K=2$ & $K=3$ & $K=4$ & $K=5$
                                      & $K=6$ & $K=7$\\
\hline
\hline
 
$\left(\frac{m_{soft}}{\MK}\right)^{\frac{1}{K+1}}$
&
& $2\times 10^{-8}$
& $7\times 10^{-6}$
& $1\times 10^{-4}$
& $8\times 10^{-4}$
& $3\times 10^{-3}$
& $6\times 10^{-3}$
& $1\times 10^{-2}$
      \\
\hline
$\langle S\rangle$ (GeV)
& 
& $5\times 10^{9}$
& $2\times 10^{12}$
& $4\times 10^{13}$
& $2\times 10^{14}$
& $8\times 10^{14}$
& $2\times 10^{15}$
& $3\times 10^{15}$
      \\
\hline
& $K-1$
& $5\times 10^{7}$
& $1\times 10^{5}$
& $7\times 10^{3}$
& $1\times 10^{3}$
& $400$
& $200$
& $90$
      \\
$\frac{\mu_{eff}}{m_{soft}}$ 
& $K$
& 1
& 1
& 1
& 1
& 1
& 1
& 1
      \\
 
& $K+1$
& $2\times 10^{-8}$
& $7\times 10^{-6}$
& $1\times 10^{-4}$
& $8\times 10^{-4}$
& $3\times 10^{-3}$
& $6\times 10^{-3}$
& $1\times 10^{-2}$
      \\
\hline
& $0$
& 1
& 1
& 1
& 1
& 1
& 1
& 1
      \\
 
& $1$
& $2\times 10^{-8}$
& $7\times 10^{-6}$
& $1\times 10^{-4}$
& $8\times 10^{-4}$
& $3\times 10^{-3}$
& $6\times 10^{-3}$
& $1\times 10^{-2}$
      \\
& $2$
& $3\times 10^{-16}$
& $5\times 10^{-11}$
& $2\times 10^{-8}$
& $6\times 10^{-7}$
& $7\times 10^{-6}$
& $4\times 10^{-5}$
& $1\times 10^{-4}$
      \\
$\frac{m_{Q,L}}{\langle H_i\rangle} $ 
& $3$
& $6\times 10^{-24}$
& $3\times 10^{-16}$
& $2\times 10^{-12}$
& $5\times 10^{-10}$
& $2\times 10^{-8}$
& $2\times 10^{-7}$
& $2\times 10^{-6}$
      \\
& $4$
& $1\times 10^{-31}$
& $2\times 10^{-21}$
& $3\times 10^{-16}$
& $4\times 10^{-13}$
& $5\times 10^{-11}$
& $1\times 10^{-9}$
& $2\times 10^{-8}$
      \\
& $5$
& $2\times 10^{-39}$
& $2\times 10^{-26}$
& $5\times 10^{-20}$
& $3\times 10^{-16}$
& $1\times 10^{-13}$
& $9\times 10^{-12}$
& $2\times 10^{-10}$
      \\
\end{tabular}
\caption{Non-Renormalizable
MSSM mass terms via $\langle S \rangle$. For
$m_{soft}\sim 100$ GeV, $\MK\sim 3\times 10^{17}$ GeV.}
\label{masstable2}
\end{table}
\hfill\vfill\eject

\def\NPB#1#2#3{{\it Nucl.\ Phys.}\/ {\bf B#1} (#2) #3}
\def\NPBPS#1#2#3{{\it Nucl.\ Phys.}\/ {{\bf B} (Proc.\ Suppl.) {\bf #1}}
 (19#2) #3}
\def\PLB#1#2#3{{\it Phys.\ Lett.}\/ {\bf B#1} (#2) #3}
\def\PRD#1#2#3{{\it Phys.\ Rev.}\/ {\bf D#1} (#2) #3}
\def\PRL#1#2#3{{\it Phys.\ Rev.\ Lett.}\/ {\bf #1} (#2) #3}
\def\PRT#1#2#3{{\it Phys.\ Rep.}\/ {\bf#1} (#2) #3}
\def\MODA#1#2#3{{\it Mod.\ Phys.\ Lett.}\/ {\bf A#1} (#2) #3}
\def\IJMP#1#2#3{{\it Int.\ J.\ Mod.\ Phys.}\/ {\bf A#1} (#2) #3}
\def\nuvc#1#2#3{{\it Nuovo Cimento}\/ {\bf #1A} (#2) #3}
\def\RPP#1#2#3{{\it Rept.\ Prog.\ Phys.}\/ {\bf #1} (#2) #3}
\def\etal{{\it et.\ al\/}}

\def\upenngrpa{G.\ Cleaver, M.\ Cveti\v c, L.\ Everett, J.R.\ Espinosa, 
     and P.\ Langacker} 

\def\upenngrpb{G.\ Cleaver, M.\ Cveti\v c, L.\ Everett, J.R.\ Espinosa, 
     P.\ Langacker, and J.\ Wang} 

\def\bg{\bibitem}

\end{document}